# Fast readout for large scale spin-based qubits

X. Luo, B. Bertrand, H. Niebojewski, F. Martins, C. Smith, T.-Y. Yang

*Abstract*—In this letter, we present fast readout of Pauli spin blockade phenomena and interdot coupling tunability in a silicon double quantum dot (DQD) fabricated using industry-compatible processes. The interdot couplings are tuned with a second self-aligned gate layer. The charge sensing and spin readout are performed by using gate-based reflectometry techniques. The results pave the way for scalable fast readout of large-scale industry-standard manufactured Si spin qubit arrays.

*Index Terms*—metal-oxide semiconductor (MOS), quantum dot (QD), cryogenic temperature, reflectometry, charge sensing, Pauli spin blockade (PSB), spin relaxation, quantum information processing (QIP).

## I. Introduction

Quantum computing with error correction necessitates the large-scale integration of a substantial number of qubits. A significant advantage of silicon spin qubits lies in (1) their compatibility with the existing semiconductor manufacturing, and (2) long coherence times due to low hyperfine interactions and small spin-orbit interactions. However, scaling up a spin qubit system is challenging for (1) microstructure fabrication [1-4], (2) wiring at a large scale, and (3) thermal budget in a cryogenic system.

In this work, we present the characterization of a quantum dot array fabricated by CEA Leti FDSOI line using deep UV (DUV) lithography without the need of e-beam lithography. It is noteworthy that these devices have a second gate layer as the coupling gates (J-gates), in addition to the plunger gate layer [2]. These J-gates enable the tunability of the interdot couplings, which is crucial for two-qubit gate operations [5]. To reduce the footprint, e.g. the need of an adjacent single electron transistor charge sensors and reservoir leads for injecting charges to each quantum dot, we apply gate-based reflectometry for fast sensing of Pauli spin blockade (PSB) in a double quantum dot, which is suitable for high temperature qubit operation [6-8].

## II. Device fabrication and Experiment

Figure 1 shows the typical device microstructures of the devices reported in this work. Active regions of the devices are obtained starting from a 300 mm silicon-on-insulator (SOI) wafer patterned in a MESA isolation integration. A TiN/polySi gate stack is deposited and patterned in a self-aligned scheme, enabling the formation of 80 nm pitch front gates (FGs) on two individual active Si nanowires as shown in Fig. 1(a). Reservoirs are made through in-situ doped raised epitaxy followed by NiPt silicidation for the source, drain, and gate regions. After the contacts fabrication, metallic trenches are etched to form 80 nm pitch coupling gates (JGs) that sit in-between FGs. Longitudinal and transverse cross-section TEM images of the dual-active split-gate device are shown in Figs. 1(b) and (c), respectively.

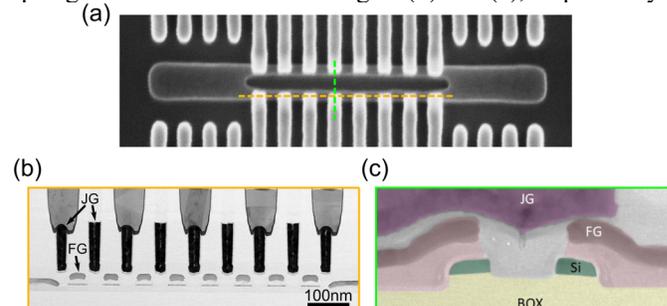

Fig. 1. (a) Top-view scanning electron microscopy of a self-aligned gate-active patterning to obtain face-to-face front gates on a dual active Si nanowire. (b) and (c) Transmission electron microscopy crosse-section images along the yellow and green dashed lines in (a), respectively. In this sample, 80 nm pitch front and exchange gates are intertwined, leading to 40 nm effective control pitch.

A typical 2×4 split-gate device structure is illustrated in Fig. 2. These devices are designed to facilitate the formation and control of a 2×N quantum dot array. The p-channel metal-oxide semiconductor (pMOS) devices feature a trench in the Si channel to form a dual quantum dot array and avoid charge movement in the transverse direction. The primary benefit of using pMOS devices is their stronger spin-orbit interactions of holes, which enable more scalable spin manipulations using electric dipole spin resonance (EDSR) [9, 10] and eliminating the need of additional components, such as micromagnets or electron spin resonance (ESR) lines [1, 10]. Specifically, this EDSR mechanism exploits the strong spin-orbit coupling inherent to valence-band hole states, which manifests as a highly anisotropic and spatially dependent g-matrix [11, 12]. When a microwave electric field is applied, it periodically displaces the hole wavefunction and modulates the confinement potential. This induces a variation in the g-tensor, acting as an effective oscillating magnetic field that coherently drives the spin transitions [13]. Crucially, the cross-sectional geometry of the device structurally enhances the magnitude of this g-factor modulation. Within our Si nanowire split-gate architecture, the

This paragraph of the first footnote will contain the date on which you submitted your paper for review.
X. Luo is with University of Cambridge, Cambridge CB3 0US, UK.
F. Martins was with Hitachi Cambridge Laboratory, Hitachi Europe Ltd. Cambridge CB3 0US, UK, and now with QuantrolOx Ltd., Oxford Centre for Innovation, New Road, Oxford, Oxfordshire OX1 1BY.

C. Smith, and T.-Y. Yang are with Hitachi Cambridge Laboratory, Hitachi Europe Ltd. Cambridge CB3 0US, UK.
B. Bertrand, H. Niebojewski are with CEA-Leti, Univ. Grenoble Alpes, F-38000 Grenoble, France

concentrated electric field at the uppermost edges of the channel forces charge carriers to accumulate and form "corner dots" [12, 14]. Because this asymmetric confinement is rigidly defined by the physical, the spatial symmetry-breaking required for efficient spin-orbit coupling is highly robust. The resulting spin-orbit interaction strength and EDSR driving efficiency are intrinsically uniform across the entire array, supporting the viability of this architecture for large-scale integration.

Gate-based reflectometry is used for the charge and spin sensing at cryogenic temperatures. The experiment setup is shown in Fig. 2. Reflectometry technique utilizes the fact that any charge transitions in the quantum dots, either an interdot charge transition (ICT) or a dot-to-lead transition change the total capacitance [15-17]. Connecting an LC resonator to a plunger gate allows for detection of such a capacitance change, thereby enabling the readout of the quantum dot's charge state. The reflectometry sensing is performed in a dry dilution refrigerator at a base temperature of 80 mK. The dc voltages are applied via a QDevil QDAC-II and the reflectometry sensing are carried out with a Quantum Machines OPX+ system for a lock-in measurement. By leveraging this setup, we achieve an integration time of 40 µs, which is by one to three orders of magnitude faster than conventional dc transport measurements [18, 19]. Consequently, this rapid reflectometry approach is highly advantageous for meeting the operational speed requirements of scalable quantum computing architectures.

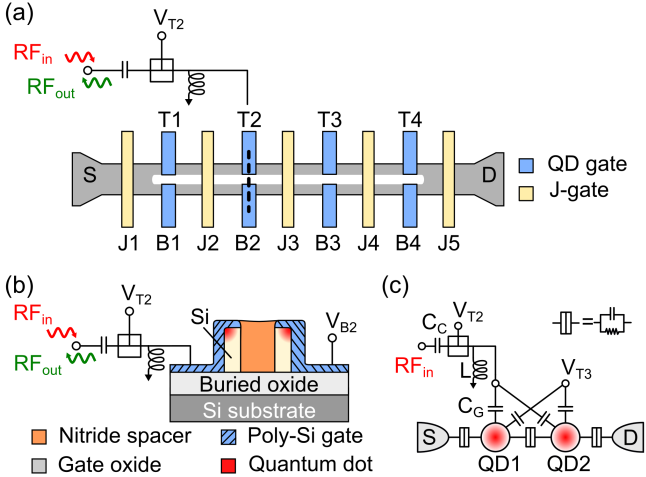

Fig. 2. Schematic top-view of a silicon transistor consisting of eight quantum dot plunger gates (T- and B-gates) and five coupling gates (J-gates). An LC resonator is connected to gate T2 for reflectometry sensing. (b) Cross-section schematic along the dashed line in (a). (c) Circuit diagram of a double quantum dot and reflectometry sensor. $C_C$ and $C_G$ are the LC resonator coupling capacitance and gate capacitance, respectively. $L$ is the resonator inductance.

### III. TRANSPORT CHARACTERIZATION

Figs. 3(a)–(b) show drain current $I_{DS}$ as a function of gate voltage $V_G$ measurements for a pMOS transistor. A small variation in the threshold voltage is essential for quantum computing multiplexing [20]. However, a moderate threshold voltage variation in our devices, -0.763±0.117 V in 56 measured gates, is observed and not preferred and requires further improvement. The stable subthreshold swing, 101.9±9.2 mV/dec, implies consistent interface-trap densities and/or gate

lever arms in the DUTs [21]. While this subthreshold swing is larger than that of conventional logic MOSFETs, we attribute the degradation primarily to the presence of interface traps. Based on the 6 nm oxide thickness of our gate stack, we estimate an interface trap density of approximately $10^{12}$ eV$^{-1}$cm$^{-2}$. This value is consistent with reported ranges for silicon spin qubit platforms [22, 23]. Given this consistent baseline interface quality, the observed threshold voltage variance is likely caused by the dopant diffusion from the ohmic contacts, resulting the lower threshold voltages in the outer gates. A similar trend has also been reported in Ref. [24].

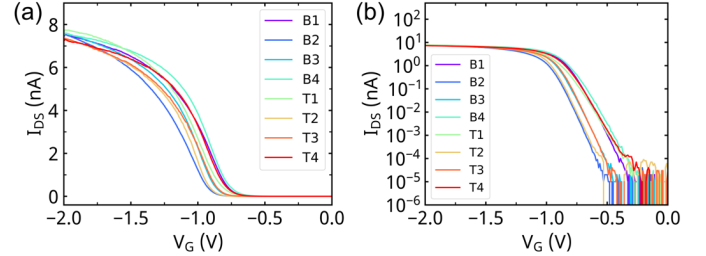

Fig. 3. Room temperature dc transport characterization of a Si nanowire transistor. (a) Measurements of drain current $I_{DS}$ as a function of gate voltage $V_G$ for eight QD plunger gates. (b) $I_{DS}$ in logarithmic scale.

### IV. REFLECTOMETRY SPIN DETECTION

In the following measurements, we focus exclusively on the DQD system formed under gates T1 and T2. To isolate this two-qubit base unit, gates T3 and T4 are held at a voltage of -2 V, preventing the formation of active quantum dots beneath them.

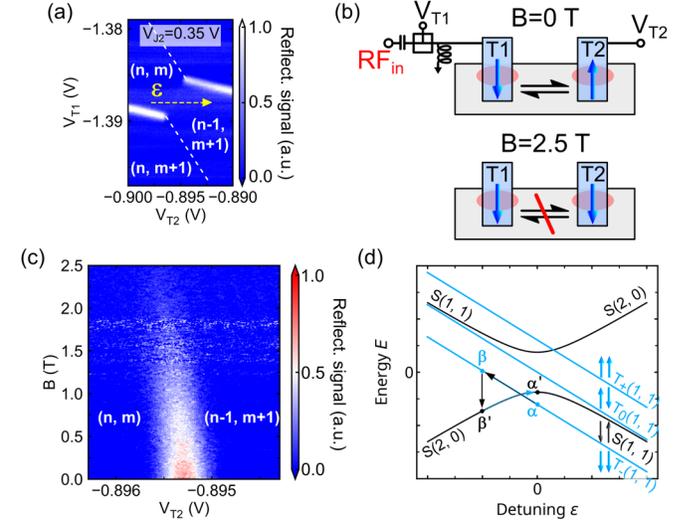

Fig. 4. (a) Gate-based reflectometry measurement as a function of gate voltages $V_{T1}$ vs $V_{T2}$ at coupling gate $V_{J2}$=0.35V and magnetic field $B$=2.5 T. $n$ ($m$) represents the number of charges in the quantum dot under gate T2 (T1). (b) Schematics of Pauli spin blockade in a double quantum dot measurement. Upper panel: at $B$=0, the spin orientations in both quantum dots are antiparallel. The charge movement can be detected using an LC resonator connected to gate T1. Lower panel: the spin orientations in each quantum dot are parallel at $B$=2.5 T, and the charge movement is forbidden. (c) Magnetospectroscopy measurement of the ICT along the detuning axis $\varepsilon$ in (a). (d) Energy $E$ as a function of detuning $\varepsilon$ diagram for singlet (S) and triplet (T) states.

Gate-based reflectometry is applied for sensing the spin-dependent states in a DQD. The interdot quantum capacitance $C_Q$, i.e. the curvature of the eigenstate energy $E$ as a function of detuning $\varepsilon$, is proportional to $\partial^2 E/\partial^2 \varepsilon$. When the spin state is in a singlet state $|\uparrow\downarrow\rangle$ or $|\downarrow\uparrow\rangle$ (Fig. 4(b) upper panel), a non-zero $C_Q$ at $\varepsilon=0$ in Fig. 4(d) can be sensed in the reflectometry signal. Figure 4(a) shows a charge occupancy $(n, m)$-$(n-1, m+1)$ (equivalent to $(2, 0)$-$(1, 1)$) interdot transition (ICT) measurement at a magnetic field $B=2.5$ T, where the spins in the DQD are in a parallel configuration, as shown in Fig. 4(b) bottom panel, resulting in a triplet state T-$|\downarrow\downarrow\rangle$ as the ground state. Consequently, Pauli spin blockade (PSB) occurs, and the measured ICT reflectometry signal in Fig. 4(a) becomes suppressed as $\partial^2 E/\partial^2 \varepsilon=0$ [16].

Figure 4(c) shows a magnetospectroscopy measurement of an ICT, where PSB appears gradually with increasing $B$-field. The asymmetric disappearance of the ICT can be explained by the fact that as the $B$-field increases, the T- state progressively lowers in energy due to Zeeman splitting. As a result, the detuning range where the T- state becomes the ground state expands asymmetrically, as illustrated in Fig. 4(d).

triplet state $\beta$ (see Fig. 4(d)), the system then quickly relaxes from state $\beta$ to singlet state $\beta'$. When the system is sequentially pulsed back to zero detuning at singlet state $\alpha'$, the system relaxes to the initial ground state $\alpha$. The recovered ICT signal can be measurable when a wait time ($t_{wait}$) between two voltage pulses is shorter than the singlet→triplet relaxation time ($T_1$). Figure 6(d) shows the reflectometry signal as a function of $t_{wait}$. The singlet population decays exponentially with $t_{wait}$, and the characteristic time $T_1$ of ~590 ns is extracted from the experimental data. It is important to note that this $T_1$ value is measured at zero detuning ($\varepsilon=0$). At this ICT, the charge states strongly hybridize. The charge admixture makes the hybridized spin states highly susceptible to electrical charge noise and phonon-mediated relaxation, creating a well-known relaxation hotspot. Consequently, the singlet-triplet relaxation rate at the charge anti-crossing is fundamentally accelerated compared to spin states isolated deep in the Coulomb blockade regime. Our measured $T_1$ time is comparable to, and even exceeds, similar measurements performed directly at the charge transition using radio-frequency reflectometry, where the relaxation times on the order of 100 ns are routinely reported in silicon quantum dot systems [25-27].

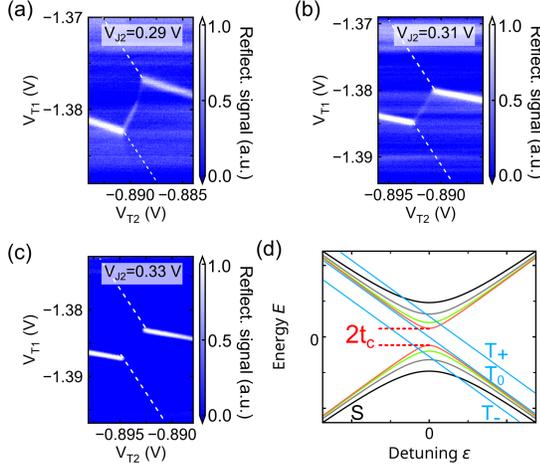

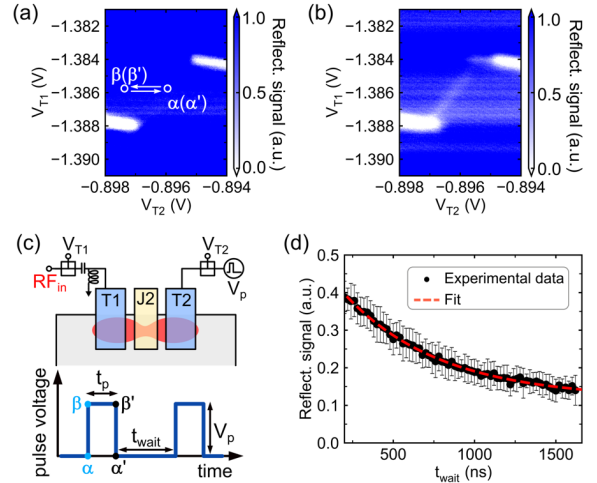

Fig. 5. ICT measurements at various coupling gate voltages $V_{J2}$=(a) 0.29 V, (b) 0.31 V, and (c) 0.33 V at $B=2.5$ T. (d) Energy $E$ vs detuning $\varepsilon$ for a singlet-triplet pair with various interdot coupling $t_c$ under a finite magnetic field.

Fig. 6. Singlet-triplet relaxation time ($T_1$) measurement. ICT measurement (a) without a pulse train and (b) with a pulse train having wait time $t_{wait}$=160 ns. (c) Upper panel: schematic of the device setup in this measurement. An LC resonator is connected to gate T1, and a pulse train is applied to gate T2 via a bias tee. Bottom panel: schematic of a pulse train. $t_p$, $t_{wait}$ and $V_p$ are the pulse length, wait time, and pulse amplitude, respectively. $\alpha$, $\alpha'$, $\beta$, and $\beta'$ correspond to the locations marked in (a). (d) Reflectometry signal as a function of $t_{wait}$. The red dashed line is an exponential decay function fit to the experimental data. In this experiment, the coupling gate voltage $V_{J2}$=0.35 V.

We also observed that by adjusting the DQD interdot coupling, the induction of PSB can be tuned. Figures 5(a)–(c) show that PSB is lifted by reducing $V_{J2}$ from 0.33 V to 0.29 V. Figure 5(d) provides an illustration of this process. The interdot coupling $t_c$ increases as the J-gate voltage reduces. From an energy perspective, this increase in $t_c$ causes the singlet state to become progressively flatter and lower. Eventually, the T- state is no longer the ground state at $\varepsilon=0$, as a result the ICT becomes measurable, i.e. the singlet state is the ground state.

Having established static control over the two-hole energy landscape and identified the precise electrostatic conditions required to isolate the Pauli spin blockade, we next turn to the dynamic properties of the system. A singlet-triplet relaxation time measurement is shown in Fig. 6. By applying a pulse train to a PSB ICT via gate T2, the ICT signal reappears in Fig. 6(b). When T- state is the ground state and the system is quickly pulsed away from zero detuning, e.g. from triplet state $\alpha$ to

## V. Conclusion

In this letter, we have characterized pMOS transistors fabricated using fully industry-compatible processes. We demonstrated the fast readout of Pauli spin blockade in a double quantum dot using gate-based reflectometry. The PSB can be induced by either applying an external magnetic field or tuning the interdot coupling. The experimental results allow us to construct the energy structures of the singlet and triplet states and further conduct $T_1$ time measurements. Our results affirm the scalability and control are achievable in silicon CMOS spin qubits for quantum computing.

## ACKNOWLEDGEMENT

This research has received funding from the European Union's Horizon 2020 QLSI project.